\documentclass[aps,prl,twocolumn,superscriptaddress,showpacs,amsmath,amssymb]{revtex4}
\usepackage[usenames,dvipsnames]{color}
\usepackage{epsf}
\usepackage{bm}
\usepackage{dcolumn}
\usepackage{latexsym}
\usepackage{amsmath}
\usepackage{amsfonts}
\usepackage{amssymb}
\usepackage{graphicx}
\usepackage[active]{srcltx}
\newcommand{\be}{\begin{eqnarray}}
\newcommand{\ee}{\end{eqnarray}}

\definecolor{darkred}{rgb}{.8,0,0}

\definecolor{darkblue}{rgb}{0,0,.7}

\begin{document}
%
\title{One-parameter class of uncertainty relations based on entropy power}
%
%
\author{Petr Jizba}
\email{p.jizba@fjfi.cvut.cz}
\affiliation{FNSPE, Czech Technical
University in Prague, B\v{r}ehov\'{a} 7, 115 19 Praha 1, Czech Republic\\}
\affiliation{ITP, Freie Universit\"{a}t Berlin, Arnimallee 14
D-14195 Berlin, Germany}
\author{Yue Ma}
\affiliation{Department of Physics, Tsinghua University, Beijing 100084, P.R.China }
\author{Anthony Hayes}
\author{Jacob A. Dunningham}
\email{J.Dunningham@sussex.ac.uk}
\affiliation{Department of Physics and Astronomy, University of Sussex,  Brighton, BN1 9QH, United Kingdom\\}

\begin{abstract}
We use the concept of entropy power to derive a new one-parameter class of information-theoretic uncertainty relations for pairs of conjugate observables in an infinite-dimensional Hilbert space.
This class constitutes an infinite tower of higher-order statistics uncertainty relations,
which allows one in principle to determine the shape of the underlying information-distribution function by measuring the relevant entropy powers.
We illustrate the capability of the new class by discussing two  examples: superpositions of vacuum and squeezed states
and the Cauchy-type heavy-tailed wave function.
\end{abstract}
%

\vspace{-1mm}
\maketitle
%
%
\noindent
{\em {Introduction}.}~---~
In 1948, Shannon laid down the foundations of modern information theory~\cite{Shannon48}.
He was instrumental in pointing out that, in contrast with discrete signals or messages where information is quantified by
(Shannon's) entropy, the case with continuous variables
is less satisfactory. The continuous version of Shannon's entropy (SE) -- the so-called
differential entropy -- may take negative values~\cite{Shannon48,Jizba2004} and so does not have
the same status as its discrete-variable counterpart. To solve a range of communication theoretic
problems related to continuous cases Shannon shifted the emphasis from the differential entropy
to another object -- entropy power (EP).
The EP represents the variance of a would-be Gaussian random variable with the same differential
entropy as the random variable under investigation. EP was used by Shannon~\cite{Shannon48}
to bound the capacity of non-Gaussian additive noise channels.
Subsequent developments in information theory confirmed the central role of the EP~\cite{Bergmans,Oohama,Lapidoth}.
On the mathematical side the EP proved to be critical in proving a strong version
of the central limit theorem with convergence in relative entropy~\cite{Artstein,Barron}.

Information theory now extends far beyond the realm of communications and the same principles and
concepts can be employed in applications that include statistical physics, biological science
and quantum mechanics~\cite{Roedder:05}. In this Letter we focus on the application of
the EP to quantum-mechanical uncertainty relations (URs).
In essence, quantum-mechanical URs place fundamental limits on the
accuracy with which one is able to know the values of different physical
quantities.
In the 1920s, Kennard and independently Robertson and Schr\"{o}dinger reformulated
original Heisenberg's UR in terms of variances of the observables~\cite{Kenn1,Robertson1929,Schroedinger1}.
In 1959, Stam~\cite{Stam:59} conjectured that the EP could be used to obtain
Heisenberg's UR.  This conjecture was bolstered in~\cite{Birula} by showing
that the usual Schr\"{o}dinger--Robertson variance-based URs
(VURs)~\cite{Schroedinger1,Robertson1929} can be derived from entropic URs.
VURs are useful and widely applied but have two major restrictions:
Firstly, the product of the conjugate variances is a single number and so can only ever give partial information about the
underlying states; secondly, variances are only useful concepts for well-behaved bell-like distributions.
For heavy-tailed or multi-peaked distributions, the variances can be large or even infinite,
making VURs ill-suited or even useless.

Here we show that  Stam's UR (and VUR)  is just a member
of a one-parameter class of EP-based inequalities, all of which
stem from yet another important information measure, namely the R\'{e}nyi entropy (RE)~\cite{Renyi1970,CovThom}
and its continuous counterparts, differential RE and R\'{e}nyi entropy power (REP).
We prove that this class constitutes an infinite tower of higher-order
cumulant URs, which allows one in principle to reconstruct the underlying
information-distribution function in a process akin to quantum
state tomography~\cite{tomography} using EPs in the place of the usual measurements.
In this respect, the strategy is not to optimize parameters in the
class of URs (e.g., to find a best bound),  but instead to identify and measure
as many EPs (associated with a given quantum state) as possible.
We illustrate this point with two examples of interest.






{\em {Entropy power.}}~---~
Let $\mathcal{X}$ be a random vector in $\mathbb{R}^{D}$ with
the probability density function (PDF), $\mathcal{F}$. The differential entropy $\mathcal{H}(\mathcal{X})$ of $\mathcal{X}$ is defined
as~\cite{Shannon48}
\begin{equation}
\mathcal{H}(\mathcal{X}) \ = \ -\int_{
\mathbb{R}^{D}}\mathcal{F}({\boldsymbol{x}})\log_{2}\mathcal{F}%
({\boldsymbol{x}})\ d{\boldsymbol{x}}\,. \label{IIA1}%
\end{equation}
The discrete version of (\ref{IIA1}) is nothing but the SE~\cite{Shannon48}.
Strictly the form shown in (\ref{IIA1}) is not a proper
entropy but rather an information gain~\cite{Jizba2004,Renyi1970}.
The entropy power
$N(\mathcal{X})$ of ${\mathcal{X}}$ is the unique number
such that~\cite{Shannon48,Costa1985}
%
%
\begin{eqnarray}
{\mathcal{H}} \left( {\mathcal{X}} \right) \
= \ \mathcal{H}\left(\sqrt{N(\mathcal{X})}\cdot {\mathcal{Z}}_{G}\right) ,
\label{3.1.0bv}
\end{eqnarray}
with ${\mathcal{Z}}_{G}$ representing a Gaussian random vector with zero mean and unit covariance matrix.
%
In the case when the Shannon differential entropy is measured in
nats the entropy power takes the form~\cite{Shannon48}
\begin{eqnarray}
N(\mathcal{X})\ = \ \frac{1}{2\pi e}\exp\left( \frac{2}{D}\ \!
\mathcal{H}(\mathcal{X})\right)  \, .
\label{A.8.b}
\end{eqnarray}

Correspondingly, the differential R\'{e}nyi
entropy ${\mathcal{I}}_p(\mathcal{X})$ of $\mathcal{X}$ is defined
as~\cite{Jizba2004,Renyi1970}
\begin{equation}
{\mathcal{I}}_p(\mathcal{X}) \ = \ \frac{1}{(1-p)}\log_{2}\left(
\int_{M}d{{\boldsymbol{x}}
}\,\mathcal{F}^{p}({\boldsymbol{x}})\right)  \, , \label{3.1.0f}
\end{equation}
where the index $p\in {\mathbb{R}}^{+}$. With the help of L'H\^{o}pital's rule one can
check that for $p\rightarrow 1$ one has ${\mathcal{I}}_p(\mathcal{X})
\rightarrow \mathcal{H}(\mathcal{X})$. Similarly to ${\mathcal{H}}$, ${\mathcal{I}}_p$
is also additive for independent events~\cite{Jizba2004}.
In analogy with the case of Shannon entropy discussed above, the $p$-th R\'{e}nyi
entropy power $N_p(\mathcal{X})$ is defined as the
solution of the equation
\begin{eqnarray}
{\mathcal{I}_{p}} \left( {\mathcal{X}} \right)
\ = \ \mathcal{I}_{p}\left(\sqrt{N_p(\mathcal{X})}\cdot
{\mathcal{Z}}_{G}\right)\,
,\label{3.1.0k}
\end{eqnarray}
where ${\mathcal{Z}}_{G}$ represents a Gaussian random vector with zero mean
and unit covariance matrix.

This type of expression was studied in~\cite{JDJ,Gardner02} where it was shown
that the only  class of solution of (\ref{3.1.0k}) is
\begin{eqnarray}
N_p(\mathcal{X})
\ &=& \ \frac{1}{2\pi} p^{-p'/p}
\exp\left(\frac{2}{D} \ \!{\mathcal{I}}_p({\mathcal{X}})\right),
\label{3.1.0e}
\end{eqnarray}
with $1/p + 1/p'= 1$ and $p\in {\mathbb{R}}^{+}$.
In addition, when $p \rightarrow 1_+$ one has
$N_p(\mathcal{X}) \rightarrow N(\mathcal{X})$.
%
%
For simplicity we have taken nats as units of information.
In passing, we may observe that from (\ref{3.1.0e}) it
follows that $N_p({\sigma}\mathcal{Z}_G) = \sigma^2$, i.e. for Gaussian
processes the EP
is simply the variance $\sigma^2$. In the case where $\mathcal{Z}_G^K$
represents a random Gaussian vector of zero mean and covariance matrix ${K}_{ij}$,
then $N_p(\mathcal{Z}_G^K) = [\det({K}_{ij})]^{1/D} \equiv |{{K}}|^{1/D}$.
Importantly, since the REs are in principle measurable~\cite{Cambpel,JizbaPRE},
the associated REPs are experimentally accessible. For some recent applications of the REs in quantum theory see, e.g.,~\cite{Bacco,Muller,Coles}.


{\em {Entropy Power Uncertainty Relations.}}~---~We start with the theorem of
Beckner and Babenko~\cite{Beckner1975,Babenko1962}.\\
{\em{Beckner--Babenko Theorem}:}
Let
\[
f^{(2)}({{\boldsymbol{x}}}) = \int_{\mathbb{R}^{D}}e^{2\pi
i{{\boldsymbol{x}} }.{{\boldsymbol{y}}}}\
f^{(1)}({{\boldsymbol{y}}})\ d{{\boldsymbol{y}}}\,,
\]
then for $p \in [1,2]$
%
%
\begin{eqnarray}
\;|(p^{\prime})^{D/2}|^{1/p^{\prime}}|\!|f^{(2)}|\!|_{p^{\prime}} \
\leq\ |p^{D/2}|^{1/p}|\!|f^{(1)}|\!|_{p}\, , \label{4.1}
\end{eqnarray}
where $p$ and $p^{\prime}$ are the H\"{o}lder
conjugates and
\begin{eqnarray}
|\!|F|\!|_{p} \equiv \left(\int_{\mathbb{R}^{D}}
|F({\boldsymbol{y}})|^{p} \ d{{\boldsymbol{y}}}\right)^{1/p},
\end{eqnarray}
for any  $F \in
L^{p}({\mathbb{R}^{D}})$. Of course, the role of $f^{(1)}$ and $f^{(2)}$
may be interchanged in
the inequality  (\ref{4.1}).
%
An elementary proof can be found, e.g., in~\cite{JDJ}.
Inequality (\ref{4.1}) is saturated only for Gaussian
functions~\cite{Babenko1962,Lieb1990}.

Anticipating quantum-mechanical applications we define
$\sqrt{{\mathcal{F}}({\boldsymbol{y}})} \equiv |f({\boldsymbol{y}})|$.
After some simple algebra we recast (\ref{4.1}) in the form~\cite{JDJ}
\begin{eqnarray}
&&\mbox{\hspace{-5mm}}\left(\int_{\mathbb{R}^{D}} [
{\mathcal{F}}^{(2)}({\boldsymbol{y}})]^{(1+t)} \
d{{\boldsymbol{y}}}\right)^{1/t}\left(\int_{\mathbb{R}^{D}}
[{\mathcal{F}}^{(1)}({\boldsymbol{y}})]^{(1+r)} \
d{{\boldsymbol{y}}} \right)^{1/r} \nonumber \\
&&\mbox{\hspace{3mm}}\leq\ [2(1+t)]^D \left|t/r
\right|^{D/2r}. \label{4.2}
\end{eqnarray}
Here, $r = p/2 -1$ and $t = p'/2 -1$. Because $1/p + 1/p'= 1$ we have the
constraint $t= -r/(2r+1)$.  Since $p\in [1,2]$ one has $r\in[-1/2,0]$ and
$t\in[0,\infty)$. Taking the negative binary logarithm
of both sides of (\ref{4.2}), we obtain
\begin{eqnarray}
&&\mbox{\hspace{-11mm}}{\mathcal{I}}_{1+t}({\mathcal{F}}^{(2)}) +
{\mathcal{I}}_{1+r}({\mathcal{F}}^{(1)}) \nonumber\\
&&\mbox{\hspace{-5mm}} \geq\ \frac{1}{r}
\log_2 [2(1+r)]^{D/2} + \frac{1}{t} \log_2 [2(1+t)]^{D/2}\, . \label{4.3b}
\end{eqnarray}
In the limit $t\rightarrow 0_+$ and $r\rightarrow 0_-$ this reduces to
%
\begin{eqnarray}
{\mathcal{H}}({\mathcal{F}}^{(2)}) +
{\mathcal{H}}({\mathcal{F}}^{(1)}) \ \geq \  \log_2\left(\frac{e}{2}
\right)^{D}\, ,
\label{4.5b}
\end{eqnarray}
which is just the classical Hirschman conjecture for Shannon's differential entropies~\cite{bourret:58,Birula}. However, the
semidefiniteness of ${\mathcal{I}}_{p}(\ldots)$ makes the URs (\ref{4.3b}) impractical.
%
%
In terms of REPs we can rewrite (\ref{4.3b}) as
\begin{eqnarray}
\mbox{\hspace{-3mm}}N_{1+t}({\mathcal{F}}^{(2)})N_{1+r}({\mathcal{F}}^{(1)})  \equiv
N_{p/2}({\mathcal{X}})N_{q/2}({\mathcal{Y}})
\geq  \frac{1}{16\pi^2} ,
\label{V.60.a}
\end{eqnarray}
where $q\equiv p'$ and the REs involved are measured in bits. This
is a one-parameter family of inequalities since $p$ and $q$ are
the H\"{o}lder conjugates. In contrast to (\ref{4.3b}) the RHS of
(\ref{V.60.a}) represents a \textit{universal} lower bound independent of $t$ and $r$.
Note that when ${\mathcal{X}}$ is a random Gaussian vector, then
${\mathcal{Y}}$ is also Gaussian and (\ref{V.60.a}) reduces to
\begin{eqnarray}
|K_{{\mathcal{X}}}|^{1/D} |K_{{\mathcal{Y}}}|^{1/D} \ = \ \frac{1}{16\pi^2} \, .
\label{V.60.aaa}
\end{eqnarray}
%
The equality follows from
the saturation of the inequality (\ref{4.1}) by Gaussian functions.

By assuming that a PDF has a
finite covariance matrix $(K_{{\mathcal{X}}}){_{{ij}}}$ then important inequalities hold, namely
%
\begin{eqnarray}
N({\mathcal{X}}) \ \leq \  |K_{{\mathcal{X}}}|^{1/D} \ \leq \
\sigma^2_{{\mathcal{X}}}\, ,
\label{14.acb}
\end{eqnarray}
with equality in the first inequality if and only if  ${\mathcal{X}}$ is a
Gaussian vector, and in the second if and only if ${\mathcal{X}}$ has
covariance matrix that is proportional to the identity matrix.
The proof of (\ref{14.acb}) is based on the non-negativity of the
Kullback--Leibler divergence and can be found, e.g. in~\cite{Dembo:91,Rioul:11}.
Inequality (\ref{14.acb}) immediately gives
\begin{eqnarray}
\!\!\sigma_{{\mathcal{X}}}^2 \sigma_{{\mathcal{Y}}}^2  \geq
|K_{{\mathcal{X}}}|^{1/D} |K_{{\mathcal{Y}}}|^{1/D}
\geq \ N({\mathcal{X}})N({\mathcal{Y}})  \geq  \frac{1}{16\pi^2} \, ,
\label{V.48.ab}
\end{eqnarray}
which saturates only for Gaussian (respective white) random vectors
${\mathcal{X}}$  and ${\mathcal{Y}}$. Note, that when $(K_{{\mathcal{X}}}){_{{ij}}}$ and $(K_{{\mathcal{Y}}}){_{{ij}}}$
exist then  (\ref{V.48.ab}) automatically
implies the conventional Robertson--Schr\"{o}dinger VUR.
Since the VUR is  implied by the Shannon EPUR alone, a natural
question arises; in what sense is the general set of inequalities
(\ref{V.60.a}) more informative than the special case $r=t=0$?

{\em {Reconstruction theorem}.}~---~To aid our intuition and, furthermore, to show the conceptual underpinning for REPURs
(12)  we first note that the differential RE can be written as ($\mathbb{E}\left[\cdots \right]$ denotes the mean value)
\begin{equation}
{\mathcal{I}}_p(\mathcal{X}) \ = \ \frac{1}{(1-p)}\log_{2} \mathbb{E}\left[ 2^{(1-p)i_{{\mathcal{X}}}} \right]  \, . \label{SEc5c.1}
\end{equation}
Here $i_{{\mathcal{X}}}({\boldsymbol{x}})  \equiv -\log_2 \mathcal{F}({\boldsymbol{x}})$
is the information in ${\boldsymbol{x}}$ (with respect to the PDF $\mathcal{F}({\boldsymbol{x}})$).
From (\ref{SEc5c.1}), the differential RE can be viewed as a reparametrized version of
the {\em cumulant generating function} of the information random variable $i_{{\mathcal{X}}}({\mathcal{X}})$.
The ensuing cumulant expansion is
\begin{eqnarray}
p\mathcal{I}_{1-p}({\mathcal{X}}) \ = \  \log_2 e \sum_{n=1}^{\infty}
 \frac{\kappa_n({\mathcal{X}})}{n!} \left(\frac{p}{\log_2 e}\right)^n\! ,
\label{SEc5c.1a}
\end{eqnarray}
where $\kappa_n({\mathcal{X}}) \equiv \kappa_n(i_{{\mathcal{X}}})$ denotes the $n$-th
cumulant of $i_{{\mathcal{X}}}({\mathcal{X}})$ (in units of {\em bits}$^n$).
From (\ref{SEc5c.1a}) it follows that REPs can be written in terms of $\kappa_n$'s.
In fact, $N_p$'s of order $p>0$
uniquely determine the underlying information PDF [for the proof see Supplemental Material~\cite{SM}].
So, the REPURs of different orders provide
additional
structural constraints between  ${\mathcal{F}}^{(1)}$ and ${\mathcal{F}}^{(2)}$
which cannot be seen with the VUR or Shannon entropy UR alone.
In this connection we list some further salient results~\cite{SM}:
%
%
\\[2mm]
a) Only Gaussian PDFs saturate all REPURs. REPURs with $r=-1/2$
can be saturated with a wider class of PDFs.
b) When $\mathcal{F}({\boldsymbol{x}})$ is close to (or equimeasurable with) a Gaussian PDF
then only $N_p$'s with $p$'s
in a neighborhood of $1$ are needed. The closer the shape is to the Gaussian PDF, the
smaller neighborhood of $1$ needed.
%
c) The non-linear nature of the RE emphasizes the more probable parts of the PDF
(typically the middle parts) for R\'{e}nyi's index $p > 1$ while for $p < 1$ the
less probable parts of the PDF (typically the tails) are accentuated. So, when the accentuated parts
in $|\psi|^2$ and $|\hat{\psi}|^2$ are close to Gaussian PDF
sectors, the associated REPUR will approach its lower bound.
In the asymptotic regime when $r = -1/2$, the saturation of the REPUR means that the
peak of ${\mathcal{F}}^{(1)}$ and tails of ${\mathcal{F}}^{(2)}$ are Gaussian, though both  ${\mathcal{F}}^{(1)}$ and ${\mathcal{F}}^{(2)}$
might be non-Gaussian.


{\em {REPUR in Quantum Mechanics}.}~---~
Let us consider state vectors that are Fourier transform duals -- the most prominent example being the configuration
and momentum space wave functions. In such a case there is a reciprocal relation between $\psi(\bf{x})$ and $\hat{\psi}(\bf{p})$, namely
\begin{eqnarray}
&&\psi({{\boldsymbol{x}}})  \ = \ \int_{\mathbb{R}^D} e^{i {{\boldsymbol{p}}}\cdot {{\boldsymbol{x}}}/\hbar} \  \! \hat{\psi}({{\boldsymbol{p}}})\ \!  \frac{d{{\boldsymbol{p}}}}{(2\pi \hbar)^{D/2}}\,
. \label{V.1.a}
\end{eqnarray}
The Riesz--Fischer equality~\cite{Hardy1959} guarantees mutual normalization $|\!|{\psi}|\!|_2 = |\!|
\hat{\psi}|\!|_{2} = 1$. Let us define
\begin{eqnarray}
&&f^{(2)}({{\boldsymbol{x}}}) \ = \ (2\pi \hbar)^{D/4}\psi(\sqrt{2\pi\hbar}\ \! {{{\boldsymbol{x}}}})\, , \nonumber \\
&&f^{(1)}({{\boldsymbol{p}}}) \ = \ (2\pi \hbar)^{D/4}\hat{\psi}(\sqrt{2\pi\hbar}\ \! {{\boldsymbol{p}}}) \, .
\label{6.2.a}
\end{eqnarray}
The factor $(2\pi \hbar)^{D/4}$ ensures that the new functions are normalized (in sense of $|\!|\ldots|\!|_2$) to unity.
With these we have the same structure of the Fourier transform  as in the Beckner--Babenko theorem. Consequently we can write
the associated RE-based URs (\ref{4.3b}) in the form
\begin{eqnarray}
&&\mbox{\hspace{-8mm}}{\mathcal{I}}_{1+t}(|{\psi} |^2) +
{\mathcal{I}}_{1+r}(|\hat{\psi} |^2) \nonumber \\
&&\mbox{\hspace{-2mm}}\ge \ \ \frac{1}{r}
\log_2 \left(\frac{1+r}{\pi\hbar}\right)^{\!D/2} + \frac{1}{t} \log_2 \left(\frac{1+t}{\pi \hbar}\right)^{\!D/2}\!\!, \label{6.3}
\end{eqnarray}
where we have made use of the identity
\begin{eqnarray}
&&\mathcal{I}_{p}(|f^{(1)}|^2) \ = \ \mathcal{I}_{p}(|\hat{\psi} |^2) - \frac{D}{2} \log_2 (2\pi \hbar)\, ,
\label{22.abc}
\end{eqnarray}
(and similarly for $f^{(2)}$).
%
%
In terms of the REP we can recast (\ref{6.3}) into the form [cf. Eq.~(\ref{V.60.a})]
\begin{eqnarray}
N_{1+t}(|\psi|^2)N_{1+r}(|\hat{\psi}|^2) \ \geq \ \frac{\hbar^2}{4} \, .
\label{32abc}
\end{eqnarray}
This looks similar to the Robertson--Schr\"{o}dinger VUR, but is now a family of
relations parametrised by $t$ (or equivalently $r$)
each having the same universal lower bound $\hbar^2/4$.
It should be noted that  the familiar VUR follows directly from Shannon's entropy power UR alone since [cf. Eq.~(\ref{V.48.ab})]
\begin{eqnarray}
\sigma_x^2 \sigma_p^2 \ \geq \  N_{1}(|\psi|^2)N_{1}(|\hat{\psi}|^2) \ \geq \ \frac{\hbar^2}{4} \, .
\end{eqnarray}
In the special case of Gaussian PDFs, the whole family reduces to the single familiar coherent-state VUR
\begin{eqnarray}
\sigma_x^2 \sigma_p^2 \ = \ N_{1+t}(|\psi_G|^2)N_{1+r}(|\hat{\psi}_G|^2) \ = \
\frac{\hbar^2}{4} \, .
\end{eqnarray}
%


{\em {Applications in Quantum Mechanics.}}~---~
As a first example we consider an optical state that is pertinent to quantum metrology~\cite{Knott2015}.
It consists of a superposition of a vacuum $|0\rangle$ and a squeezed vacuum $|z_{\zeta}\rangle$ which has the form
$
|\psi_{\zeta}\rangle = \mathcal{N}\left(|0\rangle + |z_{\zeta}\rangle\right),
$
with $\mathcal{N}=1/\sqrt{2+2(\cosh\zeta)^{-1/2}}$, and
\begin{eqnarray}
|z_{\zeta}\rangle = \sum_{m=0}^{\infty}(-1)^m\frac{\sqrt{(2m)!}}{2^m
m!}\left[\frac{(\tanh\zeta)^m}{\sqrt{\cosh\zeta}}\right]|2m\rangle\, ,
\end{eqnarray}
where  $|2m\rangle$ are even-number energy eigenstates and $\zeta\in\mathbb{R}$ is the squeezing parameter.
If we rewrite $|\psi_{\zeta}\rangle$ in the basis of the eigenstates of the position and momentum quadrature operators
\begin{eqnarray}
&&\mbox{\hspace{-5mm}}\hat{X} = \sqrt{\frac{\hbar}{2\omega}}(\hat{a} + \hat{a}^\dag), \;\;\;\; \hat{P} =
-i\sqrt{\frac{\hbar\omega}{2}}(\hat{a} - \hat{a}^\dag)\, ,
\end{eqnarray}
($\omega$ is the optical frequency and $\hat{a}$ and $\hat{a}^\dag$ are respectively the photon annihilation
and creation operators), we get for the PDFs (apart from normalization $ \mathcal{N}^2$)
%
\begin{eqnarray}
&\mbox{\hspace{-7mm}}|\psi_{\zeta}|^2\! = \! \sqrt{\frac{\omega}{\pi\hbar}}
\left|\exp\left({-\frac{\omega x^2}{2\hbar}}\right)+e^{\zeta/2}
\exp\left({-\frac{\omega e^{2\zeta} x^2}{2\hbar}}\right)\right|^2\!,\nonumber \\
&\mbox{\hspace{-6mm}}|\hat{\psi}_{\zeta}|^2 \! = \! \frac{1}{\sqrt{\pi\hbar\omega}}
\left|\exp\left({-\frac{p^2}{2\hbar\omega}}\right)+e^{-\zeta/2}
\exp\left({-\frac{e^{-2\zeta} p^2}{2\hbar\omega}}\right)\right|^2\!.
\label{27.abc}
\end{eqnarray}
%
These can be used to calculate the product $N_{1+t}(x)N_{1+r}(p)$ for different values of $r$.
The result is depicted in
Fig.~1 for three different values of the squeezing parameter. What we find is that the lower bound $\hbar^2/4$ is saturated
for both $N_{\infty}(x)N_{1/2}(p)$ and $N_{1/2}(x)N_{\infty}(p)$ regardless of the squeezing
(in Fig.~1 these correspond to $r= -1/2$ and $r\to\infty$ respectively).
From our foregoing analysis of REPURs this is easy to understand
because the infinite and half indices of the EPs focus on the peak and tails of the PDF, respectively and from
(\ref{27.abc}) we see that both the $x$ and $p$ PDFs are Gaussian in the tails as well as at the peaks
(i.e., at $x=p=0$). A REPUR is saturated only when the RE-accentuated sectors in both dual PDFs are
Gaussian~\cite{SM}.
On the other hand, it is also clear that both PDFs (\ref{27.abc}) as a whole are highly non-Gaussian. We would therefore not expect
REPURs with different indices to saturate the bound. This is clearly illustrated in Fig.~1. In passing, we note
that for any $\zeta \neq 0$ the Shannon entropy power UR
is the furthest from saturating the bound, and so is
the least informative of all the family of REPURs.

By way of comparison, we can also calculate the VUR for the state $|\psi_{\zeta}\rangle$. The variances involved are
\begin{eqnarray}
\langle(\varDelta X)^2\rangle_{\zeta}\! &= \mathcal{N}^2 \frac{ \hbar}{\omega}
\left[ \frac{1}{2}(1+e^{-2\zeta}) +\sqrt{{\rm sech}\zeta}(1-\tanh\zeta) \right], \nonumber \\
\langle(\varDelta P)^2\rangle_{\zeta}\! &= \mathcal{N}^2 \hbar\omega
\left[ \frac{1}{2}(1+e^{2\zeta}) +\sqrt{{\rm sech}\zeta}(1+\tanh\zeta) \right]. \nonumber
\end{eqnarray}
For $\zeta=0$, we have $\langle(\varDelta X)^2\rangle_{0}\langle(\varDelta P)^2\rangle_{0} = \hbar^2/4$, i.e. the VUR is saturated.
This is no surprise because, in this case, the vacuum $|\psi_0 \rangle = |0\rangle$ is the usual (Glauber) coherent state.
However, as the squeezing parameter $\zeta$ is increased the product blows up rapidly, which makes the VUR
uninformative. So the set of REPURs outperform both the Shannon EPUR and the VUR
by providing more information on the structural features of $|\psi_{\zeta} \rangle$
via the related PDFs (e.g., Gaussian peaks and tails in $p$-$x$ quadratures).
\begin{figure}[h]
\begin{center}
\includegraphics[height=2.25in] {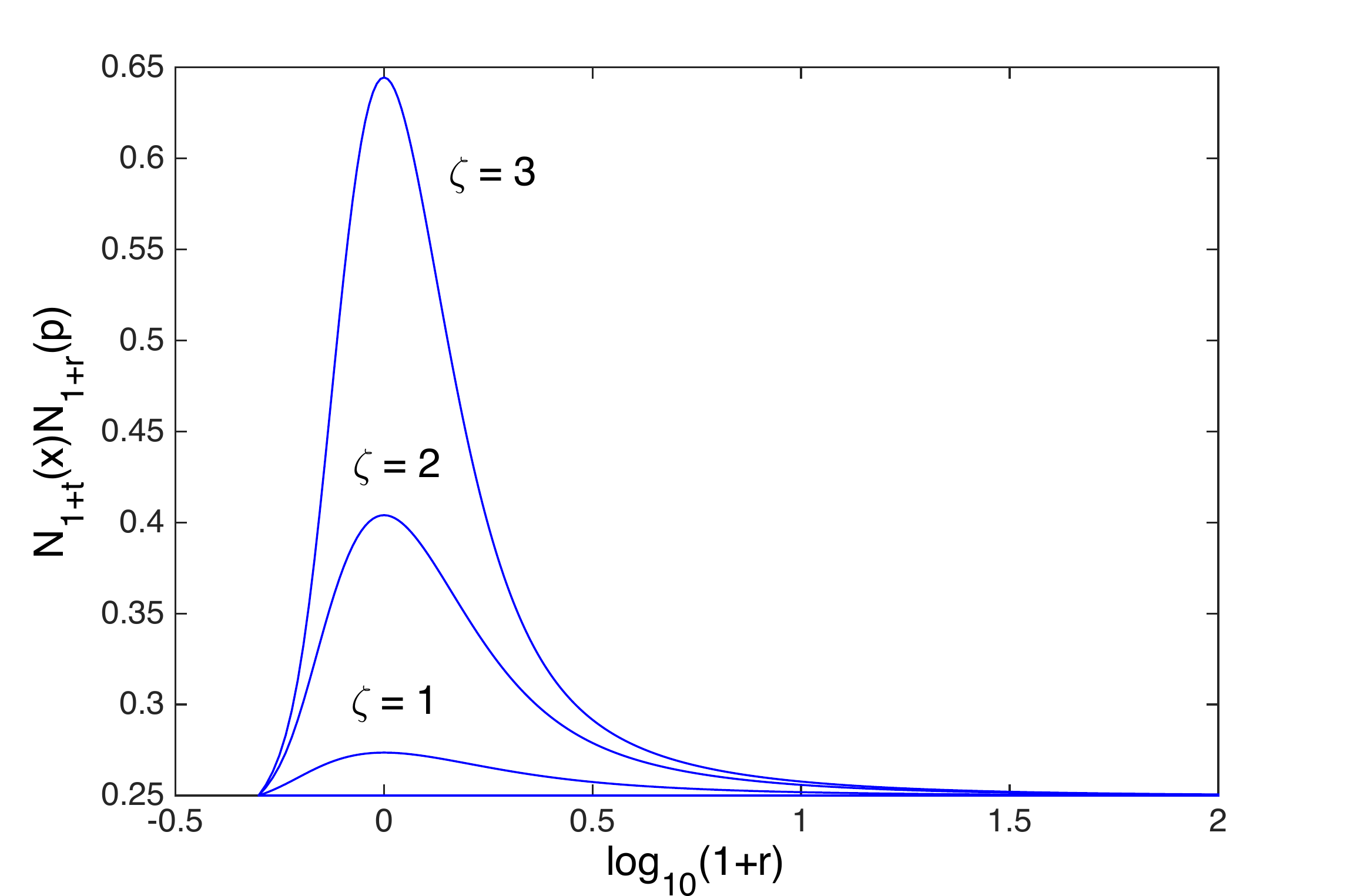} \caption{
Plot of $N_{1+t}(x)N_{1+r}(p)$ (in units of $\hbar^2$) for the state $|\psi_{\zeta} \rangle$
as a function of $\log_{10}(1+r)$  and different values of the squeezing parameter,
$\zeta$. The lower bound $\hbar^2/4$ is saturated for both $N_{\infty}(x)N_{1/2}(p)$
and $N_{1/2}(x)N_{\infty}(p)$. For other indices, REPURs deviate from the bound with the maximum
deviation at $r=0$, which corresponds to Shannon's EP.\\[-7mm]}
\label{Fig:vsv}
\end{center}
\end{figure}
Similar type of behavior can be also seen in a particular class of Schr\"{o}dinger cat  states represented by two superposed Glauber coherent states with the variable amplitude parameter~\cite{JDJ}. In the aforesaid case the Fourier transform duals were chosen to be two orthogonal phase quadratures ($x_0$ and $x_{\pi/2}$).
Specifically for $r = -1/2$ and $r \rightarrow \infty$ it was observed that the
entropic inequality (\ref{4.3b}) (and hence also the associated REPUR) were
saturated for the amplitude parameter $\beta < 1/2$, which according
to~\cite{SM} implies Gaussianity of the respective tails and peaks in state
PDFs. Since the REPUR is not saturated for $\beta \geq 1/2$ either peaks or
tails cannot be Gaussian. Closer analysis indeed revealed that the state PDF's
for $\beta \geq 1/2$  start to develop two separated peaks corresponding to the
separation of two overlapping Gaussian wave packets. In addition, for any $r$ 
the REPURs are  for large $\beta$ independent of the value of $\beta$.  This is
a consequence of two facts: a) for large $\beta$ the two Gaussian wave packets
no longer overlap and b)  REPs are immune to piecewise rearrangements of the PDF~\cite{JDJ,SM}.

We note that the conventional VUR does not pose any restriction on the
variance of the observable whose conjugate observable
has a PDF with infinite covariance matrix. So, such a state is maximally uncertain.
In contrast to this, the set of related REPURs
brings considerably more information
about the structure of these states.
To illustrate this we discuss in our second example a power-law tail wave packet (PLTWP).
PTLWPs are archetypal examples of quantum states with anomalous (scaling) behavior during their
temporal evolution~\cite{Mantegna:00}.
For definiteness we will consider the PLTWP of the form
\begin{eqnarray}
{\psi}(x) \ = \ \sqrt{\frac{\gamma}{\pi}} \ \! \sqrt{\frac{1}{\gamma^2 + (x-m)^2}}\, ,
\end{eqnarray}
which entails
the Cauchy PDF with a scale parameter $\gamma$ and median $m$. The Fourier transform
reads
\begin{eqnarray}
&&\hat{\psi}(p) \ = \ e^{-imp/\hbar}\ \! \sqrt{\frac{2\gamma}{\pi^2 \hbar }}\ \! K_0(\gamma |p|/\hbar)\, ,
\end{eqnarray}
($K_0$ is the modified Bessel function). With these results we can
immediately write two representative REPURs
\begin{eqnarray}
&&N_1(|\hat{\psi}|^2)N_1(|\psi|^2) \ = \  0.0052  \ \!\hbar^2 \pi^4 \ > \ {\hbar^2}/{4} \, ,\label{78.aaa}\\
&&N_{1/2}(|\hat{\psi}|^2)N_{\infty}(|\psi|^2) \ = \ \frac{\hbar^2}{4}\, .
\label{79.aaa}
\end{eqnarray}
Note also that $\langle(\varDelta p)^2\rangle_{\psi} = \hbar^2\pi/16c^2$ and $\langle(\varDelta x)^2\rangle_{\psi} \rightarrow \infty$
(the latter behavior is symptomatic of many PLTWPs), and so the Schr\"{o}dinger--Robertson's VUR is completely uninformative.
What can we conclude from (\ref{78.aaa})--(\ref{79.aaa})? First,
the REPUR (\ref{79.aaa}) is saturated. This implies that the peak part
of $|\psi|^2$ and the tail part of $|\hat{\psi}|^2$ are Gaussian
(as can be directly checked). Shannon's EPUR (\ref{78.aaa}) implies:
a) the involved PDFs are not Gaussian,
b) in contrast to other REPURs it quantifies only shape structures of PDFs but is $\gamma$ insensitive~\cite{SM},
c) from (\ref{4.5b}) [cf. also (\ref{22.abc})] the lower bound of Hirschman's UR is $\log_2(\pi \hbar e)$ while (\ref{78.aaa}) gives $\log_2(\pi \hbar e) + 0.5141$, so
one could still gain $0.5141$ bits of information should the system by prepared in a Gaussian state.
Finally, we note that $N_{\infty}(|\hat{\psi}|^2)=0$ and $N_{1/2}(|\psi|^2)\rightarrow \infty$, hence the related REPUR is indeterminate
(in fact, regularization dependent). This behavior is easy to understand. For a strongly leptokurtic PDF
(such as $|\psi|^2$)
$N_{1/2}$ accentuates the very flat power-law tails of $|\psi|^2$, and hence $N_{1/2}$ represents the variance of a very flat (almost equiprobable) Gaussian PDF.
Similarly, $N_{\infty}$ accentuates only the peak part of  $|\hat{\psi}|^2$ that is sharply (almost $\delta$-function) peaked,
and so  $N_{\infty}$ represents the variance of the Gaussian PDF with zero spread. Let us finally mention that in~\cite{SM} it is shown how to deduce from REPs  the scaling characteristics  for L\'{e}vy stable and Laplacian PLTWPs.

{\em {Conclusions.}}~---~
In this Letter we have formulated a new one-parameter class of R\'{e}nyi-entropy-power based
URs for pairs of observables in an infinite-dimensional Hilbert space.
The tower of inequalities obtained
possess a clear advantage over the
single VUR by revealing the finer structure of the underlying PDFs further to their standard deviations.
This was demonstrated on two relevant quantum mechanical examples and mathematically substantiated
via the reconstruction theorem. We have also
established a new formal link between the Robertson--Schr\"{o}dinger VUR
and Shannon--Hirschman UR and
highlighted the limited scope of the VUR.
Notably, we have shown that the  Robertson--Schr\"odinger VUR is a simple
consequence of the REPUR with the index $r=0$ while other REPURs in the class
set fundamental (irreducible) limits on higher order cumulants in conjugate
information PDFs.

P.J. was supported by the GA\v{C}R Grant GA14-07983S.
J.D. acknowledges support from DSTL and the UK EPSRC through the
NQIT Quantum Technology Hub (EP/M013243/1).
\vspace{-4mm}

%

\end{document}


%
\title{Supplementary Material for ``One-parameter class of uncertainty relations based on entropy power''}
%
%
\author{Petr Jizba}
\email{p.jizba@fjfi.cvut.cz}
\affiliation{FNSPE, Czech Technical
University in Prague, B\v{r}ehov\'{a} 7, 115 19 Praha 1, Czech Republic\\}
\affiliation{ITP, Freie Universit\"{a}t Berlin, Arnimallee 14
D-14195 Berlin, Germany}
\author{Yue Ma}
\affiliation{Department of Physics, Tsinghua University, Beijing 100084,
P.R.China }
\author{Anthony Hayes}
\author{Jacob A. Dunningham}
\email{J.Dunningham@sussex.ac.uk}
\affiliation{Department of Physics and Astronomy, University of Sussex,
Brighton, BN1 9QH, United Kingdom\\}
\maketitle


{\bf Note:} equations and citations that are related to the main text are shown in red.

\subsection{Reconstruction theorem\label{SEc5c}}

To show the conceptual underpinning for REPURs
{\cdred (12)} and {\cdred (22)} we begin with
the cumulant expansion {\cdred (17)}, i.e.
\begin{eqnarray}
p\mathcal{I}_{1-p}({\mathcal{X}}) \ = \  \log_2 e \sum_{n=1}^{\infty}
 \frac{\kappa_n({\mathcal{X}})}{n!} \left(\frac{p}{\log_2 e}\right)^n\! ,
\label{SEc5c.1a}
\end{eqnarray}
where $\kappa_n({\mathcal{X}}) \equiv \kappa_n(i_{{\mathcal{X}}})$ denotes the $n$-th cumulant of the information random variable $i_{{\mathcal{X}}}({\mathcal{X}})$ (in units of {\em bits}$^n$). We note that
$\kappa_1({\mathcal{X}}) =  \mathcal{H}({\mathcal{X}})$ and $\kappa_2({\mathcal{X}}) =  \mathbb{E}\left[i_{{\mathcal{X}}}({\mathcal{X}})^2\right]- (\mathbb{E}\left[i_{{\mathcal{X}}}({\mathcal{X}})\right])^2$ is the varentropy.
Now using the identity
\begin{eqnarray}
&&\mbox{\hspace{-0.9cm}}\mathcal{I}_{1-p}({\mathcal{X}}) \ = \ \mathcal{I}_{1-p}(\sqrt{N_{1-p}(\mathcal{X})}\cdot
{\mathcal{Z}}_{G})
\ = \  \frac{D}{2} \log_2 \left[
2\pi(1-p)^{-1/p} N_{1-p}(\mathcal{X})\right]
\! ,
\label{SEc5c.2a}
\end{eqnarray}
we can recast (\ref{SEc5c.1a}) as
\begin{eqnarray}
\mbox{\hspace{-0.5cm}} \log_2 \left[ N_{1-p}(\mathcal{X})\right] \ &=& \  \log_2 \left[\frac{(1-p)^{1/p}}{2\pi} \right]
\ + \  \frac{2}{D} \sum_{n=1}^{\infty} \frac{\kappa_n({\mathcal{X}})}{n!} \left(\frac{p}{\log_2 e}\right)^{n-1}\! .
\label{SEc5c.3a}
\end{eqnarray}
From (\ref{SEc5c.3a}) one can see that
\begin{eqnarray}
\kappa_n({\mathcal{X}}) \ &=& \ \left.  \frac{nD}{2}(\log_2 e)^{n-1} \frac{d^{n-1}
\log_2 \left[ N_{1-p}(\mathcal{X})\right]}{dp^{n-1}}\right|_{p=0}
\ - \  \frac{D}{2}(\log_2 e)^n \left[(n-1)! - \delta_{1n}\log 2\pi\right]\, ,
\label{SEc5c.4aa}
\end{eqnarray}
which, in terms of the Gr\"{u}nwald--Letnikov derivative formula (GLDF)~\cite{Samko}, allows us to write
\begin{eqnarray}
&&\mbox{\hspace{-0.5cm}} \kappa_n({\mathcal{X}}) \ =  \lim_{\varDelta \rightarrow 0}\frac{nD}{2}
\frac{(\log_2 e)^{n} }{\varDelta^{n-1}}\ \! \sum_{k=0}^{n-1} (-1)^k \binom{n-1}{k} \log
\left[ N_{1+k\varDelta }(\mathcal{X})\right]
\ -  \frac{D}{2}(\log_2 e)^n \left[(n-1)! - \delta_{1n}\log 2\pi\right]\, .
\label{SEc5c.5aa}
\end{eqnarray}
So, in order to determine the first $m$ cumulants of $i_{{\mathcal{X}}}({\mathcal{X}})$ we need to know all
$N_{1}, N_{1+\varDelta}, \ldots, N_{1+(m-1)\varDelta}$ entropy powers. In practice $\varDelta$ corresponds
to a characteristic resolution scale for the entropy index which is typically of the order $10^{-2}$.
Note that (\ref{SEc5c.4aa})  can be given an alternative form, namely
\begin{eqnarray}
\kappa_n({\mathcal{X}}) \ &=& \ \left.  \frac{nD}{2}(\log_2 e)^{n} \frac{d^{n-1} \log
\left[ N_{1-p}(\mathcal{X})\right]}{dp^{n-1}}\right|_{p=0} \ + \ \kappa_n({\mathcal{Z}}_{G}^{\ide})\, ,
\label{SEc5c.5aaa}
\end{eqnarray}
where $\kappa_n({\mathcal{Z}}_{G}^{\ide}) \equiv \kappa_n(i_{{\mathcal{Y}}}) $
and ${{\mathcal{Y}}}$ is the reference random variable distributed with respect
to the {\em Gaussian} distribution ${\mathcal{Z}}_{G}^{\ide}$ with the {\em unit} covariance matrix.
Analogously we can formulate (\ref{SEc5c.5aaa}) in terms of the GLDF.

When all cumulants exist then the problem of recovering the underlying PDF for $i_{{\mathcal{X}}}({\mathcal{X}})$
is equivalent to the {\em  Stieltjes} moment problem~\cite{Reed1975}. In fact, the PDF in question can be reconstructed in terms of sums involving orthogonal polynomials (e.g., the Gram--Charlier A series or the Edgeworth series~\cite{Wallaca:58}), the inverse Mellin transform~\cite{Zolotarev}
or via various maximum entropy techniques~\cite{Frontini:98}.
Pertaining to this, the theorem of Marcinkiewicz~\cite{Lukacz} implies that
there are no PDFs for which $\kappa_m = \kappa_{m+1} = \ldots = 0$ for some $m \geq 3$. In other words,
the cumulant generating function cannot be a finite-order polynomial of degree greater than $2$.
The important exceptions, and indeed the only exceptions to Marcinkiewicz's theorem, are the {\em Gaussian} PDFs which can have
the first two cumulants nontrivial and $\kappa_3 = \kappa_4 = \ldots = 0$.
Thus, apart from the special case of Gaussian PDFs where only $N_{1}$ and $N_{1+\varDelta}$ are needed,
one needs to work with as many entropy powers $N_{1+k\varDelta}, k\in {\mathbb{N}}$ (or ensuing REPURs) as possible
to receive as much information as possible about the structure of the underlying PDF. In theory, the whole infinite
tower of REPURs would be required to uniquely specify a system's information PDF.
Note that for {\em Gaussian} information PDFs  one needs only $N_{1}$ and $N_{1+\varDelta}$ to
reconstruct the PDF uniquely. From (\ref{SEc5c.3a}) and (\ref{SEc5c.5aa}) we see that knowledge of $N_1$ corresponds to
$\kappa_1({\mathcal{X}}) =  \mathcal{H}({\mathcal{X}})$ while $N_{1+\varDelta}$ further determines $\kappa_2$, i.e. the varentropy. Since $N_1$ is involved (via (\ref{SEc5c.5aa}))
in the determination of all cumulants, it is the most important entropy power in the tower.

\subsection{Information scan and ``tomography'' of $\mathcal{F}({\boldsymbol{y}})$\label{Tomography}}

\begin{figure}[ht]
	\hspace*{-1cm}
	\includegraphics[scale=0.49]{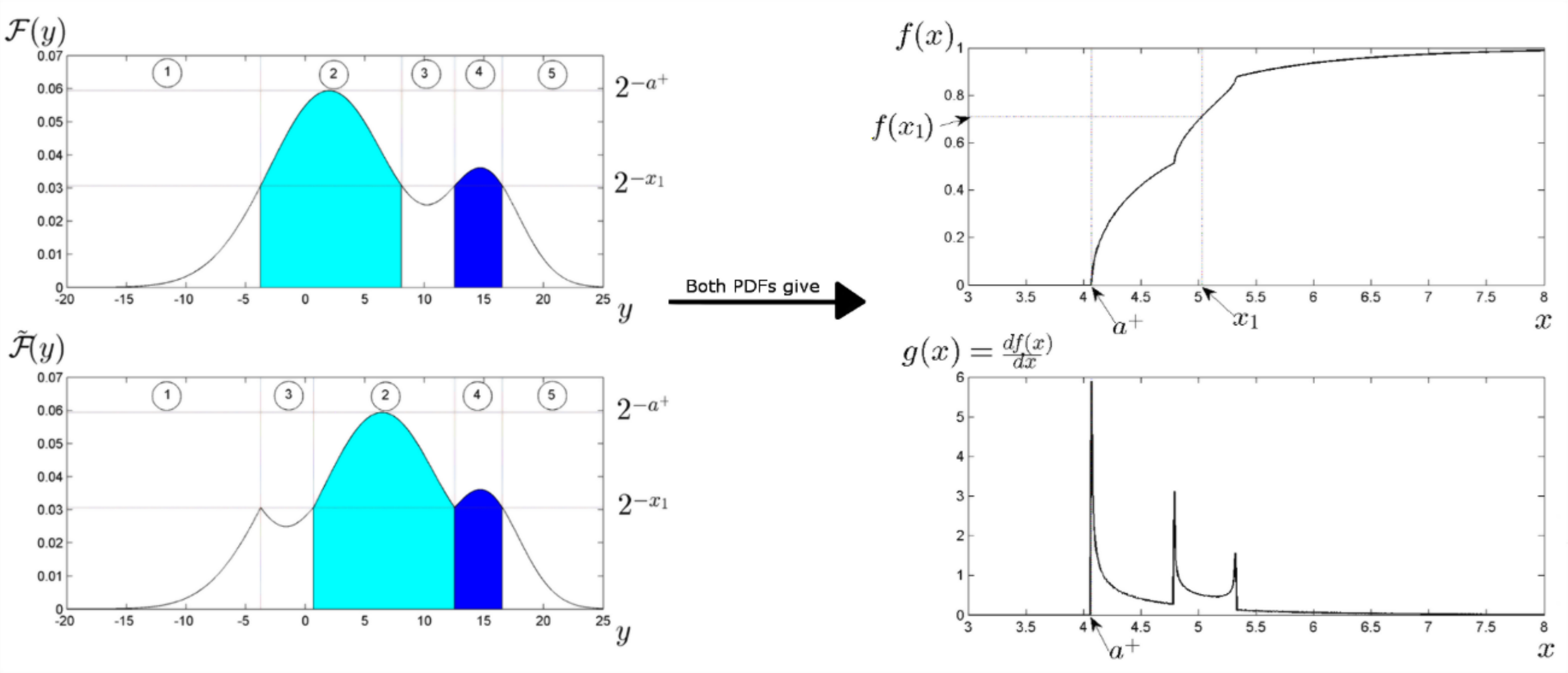}
	\caption{\footnotesize{Information scan of $\mathcal{F}(y)$. Cumulative distributions $f(x)$ measures the area of $\mathcal{F}(y)$ within the limits dictated by the intercept the line $\mathcal{F}(y) = 2^{-x}$ with $\mathcal{F}(y)$ (shaded area). For the entropy measured in nats $2^{-x} \rightarrow e^{-x}$. Information PDF $g(x)$ represents the rate of change of the area of the cumulative distribution $f(x)$. Note that $f(x)$ and $g(x)$ are identical
		for {\em equimeasurably} rearranged PDFs $\mathcal{F}({y})$ and
$\tilde{\mathcal{F}}(y)$. The number of peaks of  $g(x)$ and their positions
correspond to the {\em number} and {\em heights} of the stationary points of
both $\mathcal{F}({y})$ and $\tilde{\mathcal{F}}(y)$. The peak structure of
$g(x)$ is one of the invariant characteristics of any equimeasurable family of
PDFs.}}
	\label{fig1}
\end{figure}

\begin{figure}[ht]
	\hspace*{0.2cm}
	{\includegraphics[scale=0.455]{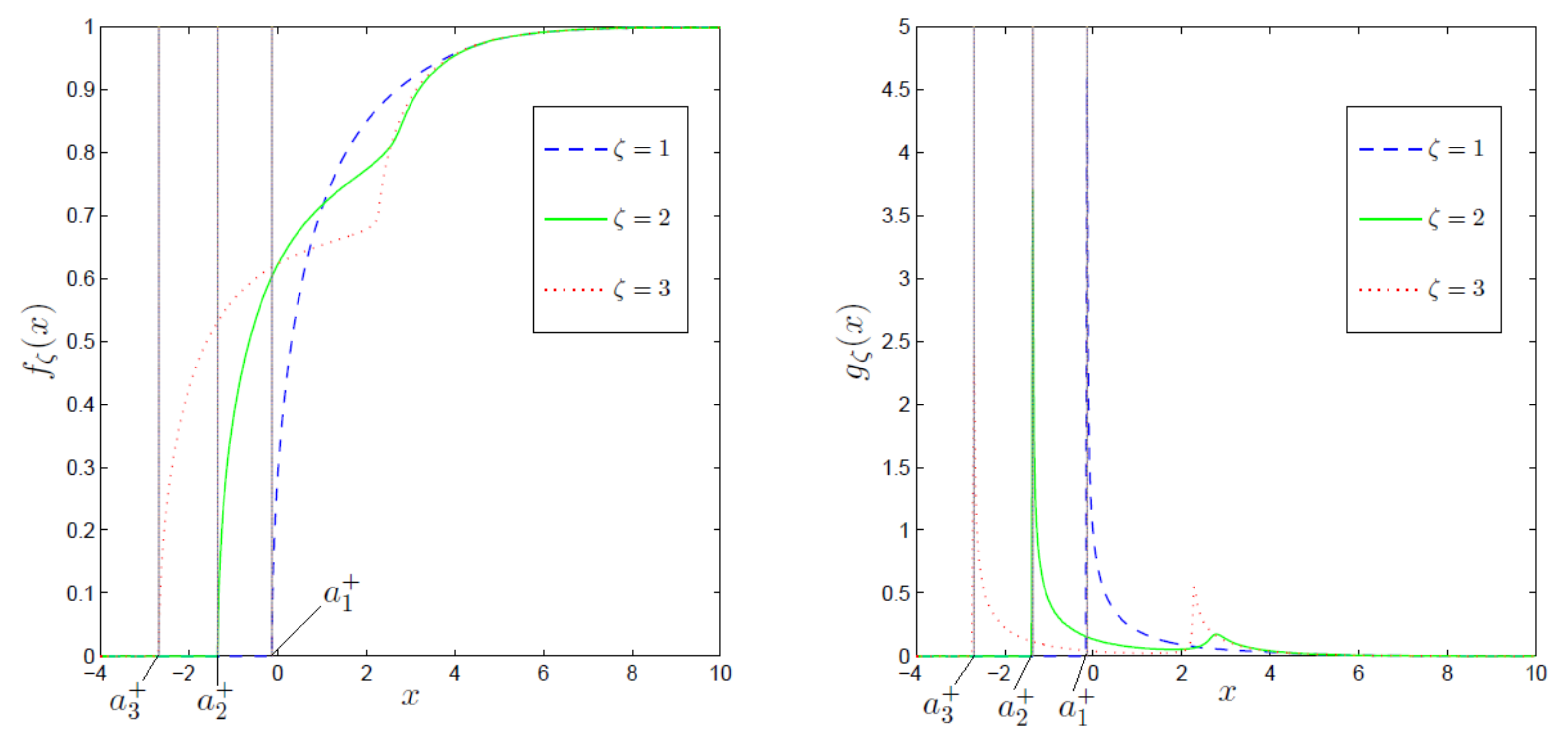}\label{a}}\\
	\vspace{5 mm}
	{\includegraphics[scale=0.455]{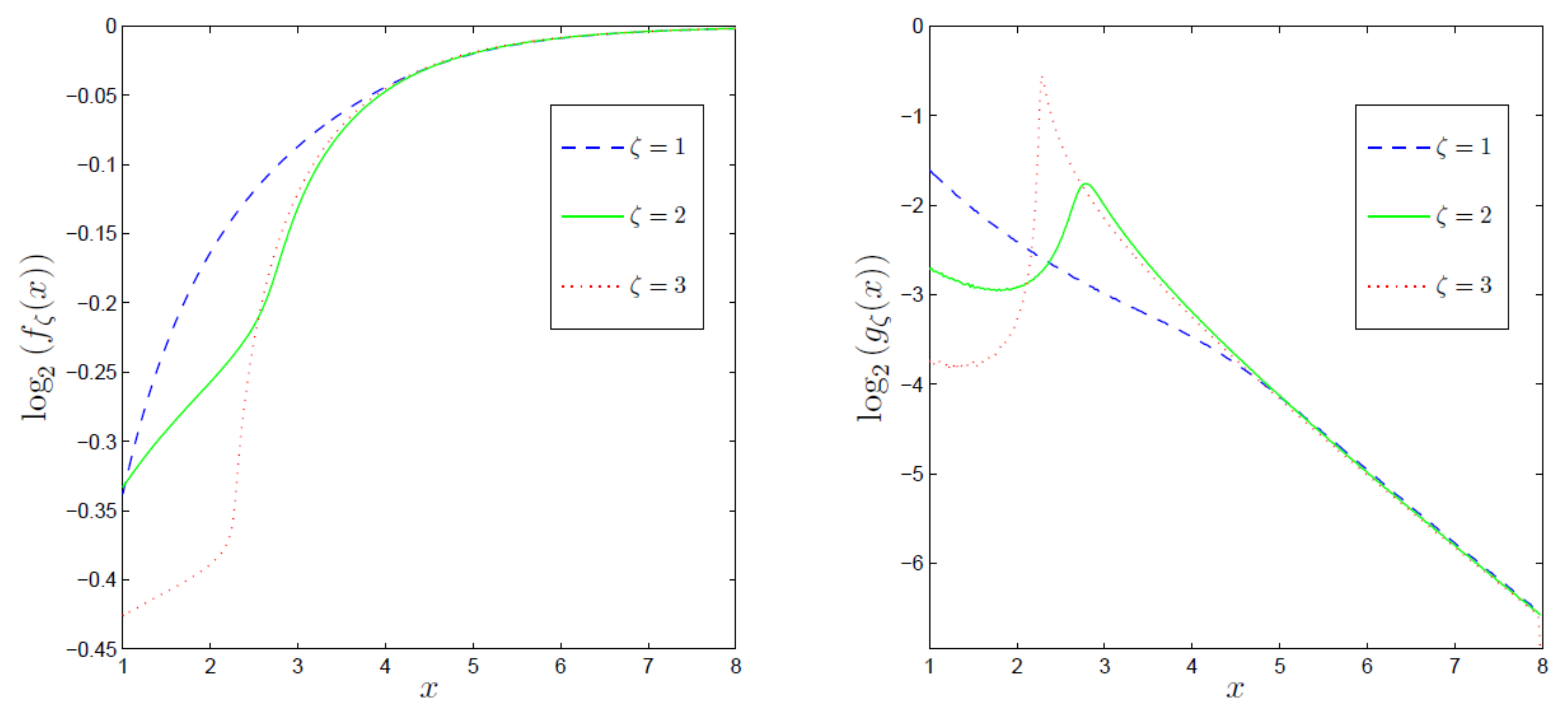}\label{b}}
\caption{\footnotesize{(a) Cumulative distributions $f_{\zeta}(x)$ and information PDFs $g_{\zeta}(x)$ for the three representative squeezed states $\psi_{\zeta}(y)$ from the main text and (b) the logarithmic scaling of $f_{\zeta}(x)$ and $g_{\zeta}(x)$ for respective values of $\zeta$ which depicts the tail behaviour (corresponding to $x>1$) of the PDF $\mathcal{F}(y) = |\psi_{\zeta}(y)|^2$. From (a) we can clearly see that the larger value of $\zeta$ the higher peak of  $\mathcal{F}({y})$. Then the peaks have heights $2^{-a^{+}_{\zeta}}$ for the respective values of $\zeta$.
For $\zeta = 2$ and $\zeta = 3$ we may observe that a new peak is formed around
$x=3$. This corresponds to an abrupt change in the shape of the PDF which
happens at the height $\mathcal{F}(y) =|\psi_{\zeta}(y)|^2 = 2^{-x}$. From (b)
we can read off the tail behavior of $|\psi_{\zeta}(y)|^2$. The best-fit
analysis reveals that $|\psi_{\zeta}(y)|^2$ has the Gaussian tail ($a=2$)
[cf. (24)].
}}
\label{fig2}
\begin{picture}(20,7)
\put(-230,250){ $b)$ } \put(-230,455){ $a)$ }
\end{picture}
\end{figure}

Let us now clarify the connection between the PDF for the random variable
$i_{{\mathcal{X}}}({\mathcal{X}})$ --- the so-called information PDF,
and the PDF for the random variable ${\mathcal{X}}$. To this end we define the cumulative
distribution function of the random variable $i_{{\mathcal{X}}}({\mathcal{X}})$, i.e.
\begin{eqnarray}
\!f(x) \ = \ \int_{-\infty}^x df(i_{{\mathcal{X}}})
 \ = \ \int_{\mathbb{R}^D} \mathcal{F}({\boldsymbol{y}}) \theta(\log_2 \mathcal{F}({\boldsymbol{y}}) + x) d{\boldsymbol{y}}\, .
 \label{SEc5c.7aa}
\end{eqnarray}
We see that $f(x)$ represents the probability that the random variable  $i_{{\mathcal{X}}}({\boldsymbol{y}}) = \log_2 1/\mathcal{F}({\boldsymbol{y}})$ is equal or less than $x$, and $ df(i_{{\mathcal{X}}})$ denotes the corresponding  probability measure.
Taking the  Laplace transform of both sides of (\ref{SEc5c.7aa}), we get
\begin{eqnarray}
\!\!\mathcal{L}\{f\}(s)  \ = \  \int_{\mathbb{R}^D}\!\!\! \mathcal{F}({\boldsymbol{y}}) \ \! \frac{e^{s\log_2 \mathcal{F}({\boldsymbol{y}}) }}{s} \ \! d{\boldsymbol{y}} \ = \ \frac{{\mathbb{E}}\left[e^{s\log_2 \mathcal{F} }  \right]}{s} \, .
\end{eqnarray}
By assuming that $f(x)$ is smooth then the PDF associated with $i_{{\mathcal{X}}}({\mathcal{X}})$ is
\begin{eqnarray}
g(x) \ = \ \frac{df(x)}{dx} \ = \  \mathcal{L}^{-1}\left\{ {\mathbb{E}}\left[e^{s\log_2 \mathcal{F}} \right]  \right\}\!(x)\, .
\label{9aa}
\end{eqnarray}
Setting $s = (p-1)\log 2$ we have
\begin{eqnarray}
\mathcal{L}\{g\}(s = (p-1)\log 2) \ = \  \mathbb{E}\left[ 2^{(1-p)i_{{\mathcal{X}}}} \right]\, ,
 \label{SEc5c.8aa}
\end{eqnarray}
or equivalently
\begin{eqnarray}
\int_{\mathbb{R}} g(x) 2^{(1-p)x}dx \ = \ \int_{\mathbb{R}^D} d{{\boldsymbol{y}}} \ \! \mathcal{F}^{p}({\boldsymbol{y}})\, .
\label{II.11aa}
\end{eqnarray}
Since $\mathcal{L}\{g\}(s)$ is the moment-generating function of the random variable $i_{{\mathcal{X}}}({\mathcal{X}})$ (with PDF $g(x)$)
one can find all moments (if they exist) by taking the derivatives of $\mathcal{L}\{g\}$ with respect to $s$. By taking
the logarithm of both sides of (\ref{SEc5c.8aa})-(\ref{II.11aa}) and comparing with {\cdred (16)}
we see that all cumulants $\kappa_n$ of $i_{{\mathcal{X}}}({\mathcal{X}})$ can be obtained by performing derivatives with respect to $s$ (or equivalently $p$). While the left-hand-side directly provides cumulants for the random variable $i_{{\mathcal{X}}}({\mathcal{X}})$ (with PDF $g(x)$),
the right-hand-side phrases these cumulants in terms of  ${\mathbb{E}}\left[(\log_2\mathcal{F})^m \right]$ (which, as we have seen, can be rephrased via EPs of the PDF $\mathcal{F}({\boldsymbol{y}})$). So, while the PDF $\mathcal{F}({\boldsymbol{y}})$ enters via its EPs,  the PDF $g(x)$ enters via its cumulants. We should stress, that the focus of the reconstruction theorem
is on cumulants $\kappa_n$ which can be directly used for a shape estimation of $g(x)$ but not $\mathcal{F}({\boldsymbol{y}})$.
However, by knowing $g(x)$ we have a complete ``information scan'' of  $\mathcal{F}({\boldsymbol{y}})$ as seen from FIG.~\ref{fig1}.
We can also see from FIG.~\ref{fig1} that the ``information scan'' is not unique, indeed two PDFs  (in our depiction $\mathcal{F}(y)$ and $\tilde{\mathcal{F}}(y)$)
that are rearrangements of each other --- {\em equimeasurable} PDFs have identical both $f(x)$ and $g(x)$.
Even though equimeasurable PDFs cannot be distinguished via their entropy powers, they can be, as a rule, easily distinguished via their respective
momentum-space PDFs and associated entropy powers. So the information scan has a tomographic flavor to it. In FIG.~\ref{fig2} we depict  $f(x)$ and $g(x)$ for the three
representative squeezed states $\psi_{\zeta}(y)$ from the main text. From the multi-peak structure of $g(x)$
one can determine the {\em number} and {\em height} of the stationary points.  These are invariant characteristics of a given family of equimeasurable PDFs.

From  (\ref{II.11aa})  we clearly see that the differential RE is indeed a reparametrized version of
the {\em cumulant generating function} of the information random variable $i_{{\mathcal{X}}}({\mathcal{X}})$. In addition,
when $\mathcal{F}^{p}$ is integrable for $p\in [1,2]$ (i.e., the admissible range of values of $p$ from the main text) then (\ref{II.11aa})
ensures that the moment-generating function for $g(x)$ PDF exists. So in particular, the moment-generating function exists when $\mathcal{F}({\boldsymbol{y}})$ represents L\'{e}vy alpha-stable distributions, including the heavy-tailed stable distributions (i.e, PDFs with the L\'{e}vy stability parameter $\alpha \in(0,2]$). The same holds for $\hat{\mathcal{F}}$ for $p'\in [2,\infty)$ (i.e., the admissible range of values of $p'$ from the main text) due to the Beckner--Babenko theorem.
\newpage

The inner workings of the  information scan can be illustrated also explicitly with the (generalized) Gram--Charlier~A expansion, though other --- often more efficient methods --- are also available~\cite{Wallaca:58}.
Let $\kappa_n$ be cumulants obtained from entropy powers and let $G({x})$ be some reference PDF whose cumulants are $\gamma_k$. The information PDF $g({x})$ can be then written as~\cite{Wallaca:58}
\begin{eqnarray}
g({x}) \ = \ \exp\left[\sum_{k=1}^{\infty}(\kappa_k - \gamma_k) (-1)^k \frac{ (d^k/dx^k)}{k!}   \right] G({x})\, .
\label{12abc}
\end{eqnarray}
With the hindsight [cf. (\ref{SEc5c.47aa})] we choose the reference PDF $G({x})$ to be a shifted gamma PDF, i.e.
\begin{eqnarray}
G({x}) \ \equiv\ {\mathcal{G}}({x}|a,\alpha,\beta)  \ = \ \frac{e^{-(x-a)/\beta}(x-a)^{\alpha -1}}{{\beta^{\alpha} \Gamma[\alpha]}},\;\;\;\;\;\; a < x < \infty, \; \beta > 0, \; \alpha > 0\, .
\end{eqnarray}
In doing so, we have implicitly assumed that the $\mathcal{F}(y)$ PDF is in the first approximation equimeasurable with the Gaussian PDF. To reach a corresponding matching we should choose (see ``Examples" section below) $a = a^+ = \log_2 (2\pi \sigma^2)/2$, $\alpha = 1/2$ and $\beta = \log_2 e$.
Using the fact that~\cite{cc}
\begin{eqnarray}
\frac{(\beta)^{k+1/2}}{k!} \frac{d^k}{dx^k} {\mathcal{G}}({x}|a,1/2,\beta) \ = \  \left(\frac{x-a}{\beta}\right)^{-k}L_k^{(-1/2-k)}\left(\frac{x-a}{\beta}\right) {\mathcal{G}}({x}|a,1/2,\beta)\, ,
\end{eqnarray}
(where $L_k^{\delta} $ is an associated Laguerre polynomial of order $k$ with parametr $\delta$) and given that
$\kappa_1 = \gamma_1 =  \alpha\beta + a = \log_2(2\pi \sigma^2 e)/2$,  and  $\gamma_k = \alpha \beta^k =  (\log_2 e)^k/2$ for $k>1$ we can write (\ref{12abc}) as
\begin{eqnarray}
\mbox{\hspace{-5mm}}g(x)   \ = \ {\mathcal{G}}({x}|a,1/2,\beta) \left[1 + \frac{(\kappa_2 -\gamma_2)}{\beta^{1/2}\left({x-a}\right)^{2} } \ L_2^{(-3/2)}\left(\frac{x-a}{\beta}\right) - \frac{(\kappa_3 -\gamma_3)}{\beta^{1/2}\left({x-a}\right)^{3} } \ L_3^{(-7/2)}\left(\frac{x-a}{\beta}\right) + \cdots\right].
\label{14aa}
\end{eqnarray}
If needed, one can use a relationship between the moments and the cumulants (Fa\`{a} di Bruno's formula~\cite{Lukacz}) to rewrite the expansion (\ref{14aa}) in the language of moments. For the Gram--Charlier A expansion various formal convergence criteria exist (see, e.g.,~\cite{Wallaca:58}). In particular, the expansion for nearly Gaussian equimeasurable PDFs ${\mathcal{F}(y)}$ converges quite rapidly and the series can be truncated fairly quickly. Since in this case one needs fewer $\kappa_k$'s in order to determine the information PDF $g(x)$, only EPs in the small neighborhood of the index $1$ will be needed. On the other hand, the further the $\mathcal{F}({y})$ is from  Gaussian (e.g., heavy-tailed PDFs) the higher the orders of $\kappa_k$ are required to determine $g(x)$, and hence a wider neighborhood of the index $1$ will be needed for EPs.

As mentioned, a given $g(x)$ characterizes a class of equimeasurable PDFs. An important universal characteristic of equimeasurable PDFs is their {\em tail behavior}.
In the following section we will illustrate how the tail behavior of $\mathcal{F}(y)$ can be directly deduced from the form of $g(x)$.

\subsection{Examples\label{example}}

An explicit example may help illuminate the reconstruction theorem. Consider $\mathcal{F}(y)$ to be a Gaussian PDF with
zero mean, variance $\sigma^2$ and $y\in{\mathbb{R}}$. The corresponding information PDF $g(x)$ is [cf. (\ref{SEc5c.7aa}) and (\ref{9aa})]
\begin{eqnarray}
\mbox{\hspace{-4mm}}g(x) \ = \  \frac{df(x)}{dx}
\ = \ \frac{2\sigma^2}{\log_2 e}\int_{\mathbb{R}}
\frac{\exp(-y^2/2\sigma^2)}{\sqrt{2\pi \sigma^2}} \ \!\frac{\left[\delta(y
-\sigma \sqrt{z(x)})
+ \delta(y+\sigma \sqrt{z(x)})\right]}{2 \sigma \sqrt{z(x)}} dy
 \ = \ \frac{2}{\log_2 e}\frac{\exp[-z(x)/2]}{\sqrt{2\pi z(x)}} \, .
\label{SEc5c.47aa}
\end{eqnarray}
Here $z(x) = {2x/\log_2 e  - \log(2\pi \sigma^2)}$. Note, that from $\theta(\log_2 \mathcal{F}(y) + x)$ we have $2x\geq \log_2 (2\pi
\sigma^2)$, and hence $\sqrt{z(x)}$ is real.
Distribution (\ref{SEc5c.47aa}) is the so-called {\em shifted gamma distribution}. For the random variable $\mathcal{Z} =
2\mathcal{X}/\log_2 e  - \log(2\pi \sigma^2)$ the PDF  (\ref{SEc5c.47aa}) represents the familiar $\chi^2$ distribution.
Cumulants of the shifted gamma distribution are well known~\cite{cc}
\begin{eqnarray}
\kappa_n(\mathcal{Z}_{G}^{\sigma}) =  \left\{ \begin{array}{ll}
\frac{1}{2}(\log_2 e) + \frac{1}{2} \log_2(2\pi \sigma^2) & \mbox{if
$n=1$}\, ,\\[3mm]
\frac{1}{2}(\log_2 e)^n \ \! \Gamma(n)& \mbox{if $n \geq 2$}\, .
\end{array}
\right.
\label{chi-2}
\end{eqnarray}
So, it is only $\kappa_1$ which contains non-trivial information about the shape of the $\mathcal{F}({{y}})$ PDF,
namely its typical scale (or standard deviation) $\sigma$. All the other cumulants just contain information about the fact that
the $\mathcal{F}({{y}})$ PDF is a generic Gaussian PDF. Should we know in advance that the underlying PDF is Gaussian,
then only  $\kappa_1$ would be needed. However, if we have no {\em a priori} knowledge about $\mathcal{F}$, we need to
consider all the cumulants.


The previous conclusion can also be arrived at from another angle.
Note that when $\mathcal{F}({{y}})$ is a {\em Gaussian} PDF with the covariance
matrix $(K_{{\mathcal{X}}})_{ij}$ then $ N_p(\mathcal{Z}_{G}^{K}) = |K_{{\mathcal{X}}}|^{1/D}$,
%
%
and hence due to (\ref{SEc5c.5aaa})  we have
\begin{eqnarray}
&&\mbox{\hspace{-0.7cm}} \frac{2}{nD}  (\log_2 e)^{-n} [\kappa_n(\mathcal{Z}_{G}^{K}) -
\kappa_n({\mathcal{Z}}_{G}^{\ide})]
\ = \ \left.  \frac{d^{n-1} \log \left[ N_{1-p}(\mathcal{Z}_{G}^{K})\right] }{dp^{n-1}}\right|_{p=0}
\ = \  \delta_{1n} \! \ \frac{1}{D}\log |K_{{\mathcal{X}}}|\, .
\label{II.10aa}
\end{eqnarray}
So, apart from the first cumulant (which represents the mean values of the two random variables $i_{{\mathcal{X}}}$
and $i_{{\mathcal{Y}}}$ with ${\mathcal{X}} \sim \mathcal{Z}_{G}^{K}$ and ${\mathcal{Y}} \sim \mathcal{Z}_{G}^{\ide}$,
respectively) all higher cumulants $\kappa_n(\mathcal{Z}_{G}^{K})$
are identical to $\kappa_n({\mathcal{Z}}_{G}^{\ide})$. This implies that non-trivial information on the PDF structure
of $i_{{\mathcal{X}}}$ is obtained only via $\kappa_1(\mathcal{Z}_{G}^{K}) =  \mathcal{H}({\mathcal{X}})$.
This elucidates why the Shannon entropy (and ensuing EP) is the most relevant
information measure (and EP) for the treatment of Gaussian PDFs.

By inserting (\ref{chi-2}) into (\ref{SEc5c.3a}) we obtain that the whole tower
of REPURs {\cdred (12)} saturates at once --- as it should be for the Gaussian
PDF. In fact, from the Beckner--Babenko theorem we know that saturation in a
single index would be indicative enough to guarantee that the underlying PDF is
Gaussian.

As a second example we take $\mathcal{F}(y)$ with $y\in{\mathbb{R}}$, and
consider $f(x)$ for $x \gg 1$. In this case we can write
[cf. (\ref{SEc5c.7aa})]
\begin{eqnarray}
1-f(x) \ = \ \int_{{\mathbb{R}}} \mathcal{F}(y) \theta(-\log_2 \mathcal{F}(y) - x) dy \ = \ \int_{\mathcal{F}(y) < 2^{-x}} \mathcal{F}(y) dy\, .
\end{eqnarray}
For very large $x$ only tail $y$'s will contribute to the integral. To
illustrate what this implies quantitatively we consider PDFs that are
equimeasurable with L\'{e}vy stable distributions, i.e., with the most important
class of heavy tailed PDFs. Their asymptotic behavior is known to be
\begin{eqnarray}
\mathcal{F}(y) \ \sim \ \left\{
                          \begin{array}{ll}
                            C_- |y|^{-(1+\alpha)}, & \hbox{as}~ y \rightarrow -\infty\, , \\
                            C_+ |y|^{-(1+\alpha)}, & \hbox{as}~ y \rightarrow \infty\, .
                          \end{array}
                        \right.
\end{eqnarray}
Here $\alpha$ is a continuous parameter over the range $0<\alpha < 2$. With
this we can write
\begin{eqnarray}
1-f(x) \ = \ \int_{(C_+ 2^x)^{1/(1+ \alpha)}}^{\infty} \mathcal{F}(y) dy \ + \ \int^{-(C_- 2^x)^{1/(1+ \alpha)}}_{-\infty} \mathcal{F}(y) dy\, ,
\end{eqnarray}
which implies that for large $x$
\begin{eqnarray}
g(x) \ = \ df(x)/dx \ = \  2^{-\alpha x/(1+\alpha)} \frac{\log 2}{(1+\alpha)}\left(C_+^{1/(1+\alpha)} + C_-^{1/(1+\alpha)}\right).
\end{eqnarray}
In particular, by plotting $g(x)$  on a logarithmic scale one can deduce both the scaling parameter $\alpha$  and the constants $C_+$ and $C_-$. By the same token, we can consider equimeasurable PDFs with tails
\begin{eqnarray}
\mathcal{F}(y) \ \sim \ \left\{
                          \begin{array}{ll}
                            D_- 2^{-\beta |y|^a}, & \hbox{as}~ y \rightarrow -\infty\, , \\
                            D_+ 2^{-\beta |y|^a}, & \hbox{as}~ y \rightarrow \infty\, .
                          \end{array}
                        \right.
\end{eqnarray}
Here $\beta > 0$ and $a > 0$ ($a=1$ and $a=2$ correspond to Laplacian and Gaussian tails respectively).  In this case we get
\begin{eqnarray}
g(x) \ = \ \frac{2^{-x}}{a  \beta^{1/a}}  \left[(x + \log_2 D_+)^{1/a -1} +   (x + \log_2 D_-)^{1/a -1}\right]\, .
\end{eqnarray}
So again from the tail behavior of $g(x)$ in the logarithmic scale one can deduce the tail characteristics of $\mathcal{F}(y)$.
In particular by setting $D_{\pm} = 1/\sqrt{2\pi \sigma^2}$, $\beta = 1/(2\sigma^2)$ and $a=2$ we regain the exact result (\ref{SEc5c.47aa}).
Note also that the faster the PDF $\mathcal{F}(y)$ decays for large $|y|$ the faster the information PDF $g(x)$  decays for large $x$.

\begin{figure}[ht]
	\hspace*{-1cm}
	\includegraphics[scale=0.35]{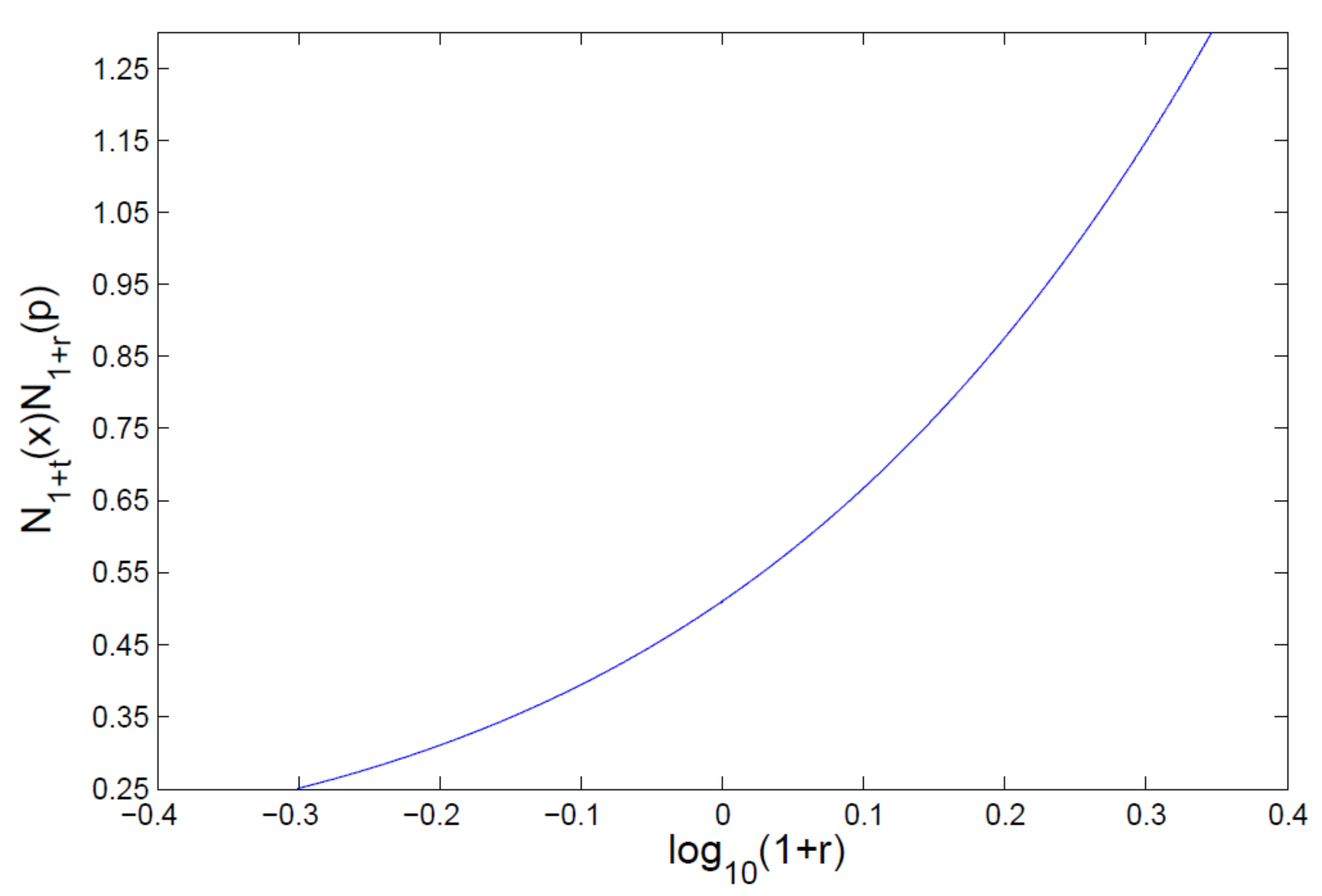}
	\caption{\footnotesize{Plot of $N_{1+t}(x)N_{1+r}(p)$ (in units of
$\hbar^{2}$) as a function of $\log_{10}(1+r)$ for the PTLWP that entails the
Cauchy PDF with $\gamma=1$ and $m=0$. As predicted by equation {\cdred (30)} in
the main text, the lower bound of $\hbar^{2}/4$ is saturated for
$N_{\infty}(x)N_{1/2}(p)$ and deviates for other indices with the notable case
of the Shannon entropy $N_{1}(x)N_{1}(p)=0.0052\hbar^{2}\pi^{4}>\hbar^{2}/4$
corresponding to $r=0$. Our numerical simulations indicate that all REPURs are
independent of both $m$ and $\gamma$, even though $N_{1+t}(x)$ and $N_{1+r}(p)$
are separately $\gamma$ dependent. This implies that REPURs quantify only
shape structures of PDFs involved --- as would be expected from the PDF without
a fundamental scale (i.e., variance). }}
	\label{fig3}
\end{figure}

\subsection{Anomalous behavior of REPURs at $r=-1/2$\label{asymptotes}}

Finally, in this section we derive a number of properties related to the anomalous behavior of the REPUR $N_{1/2}(\hat{\mathcal{F}})N_{\infty}( \mathcal{F})$ (and by symmetry also for $\hat{\mathcal{F}} \leftrightarrow \mathcal{F}$).  \\[1mm]

$a)$ For the case $t \rightarrow \infty$ we can rewrite the corresponding entropy power (with entropy measured in {\em nats}) as
\begin{eqnarray}
N_{\infty/2}(\mathcal{F})\ = \ \frac{1}{2\pi}\exp\left(-\frac{4}{D} \log|\!|f^{(2)}|\!|_{\infty}\right).
\label{tinfty1}
\end{eqnarray}
Here $|\!|f^{(2)}|\!|_{\infty} = \max_y |f^{(2)}(y)|$.  In the following we will simply write $N_{\infty}$ instead of $N_{\infty/2}$.
Note that the LHS of (\ref{II.11aa}) has the large $t$ expansion
\begin{eqnarray}
\int_{\mathbb{R}} g(x) 2^{-(1+2t)x}dx \ = \ -\frac{\log_2 e}{1+2t} \int_{\mathbb{R}} \left(\frac{d}{dx}  2^{-(1+2t)x}  \right) g(x) dx
\ = \ \frac{\log_2 e}{1+2t} \ \! g(a^+) 2^{-(1+2t)a^+} \ + \ \mathcal{O}(t^{-2})\, , \label{II.15.a}
\end{eqnarray}
where $a_+$ is the {\em onset point} at which $g$ starts to be non-zero.
By taking the $1/[2(1+t)]$ power of  (\ref{II.15.a})  and equating it
[via equation (\ref{II.11aa})] with  $|\!|f^{(2)}|\!|_{\infty}$  we obtain,
in the large $t$ limit the identity $2^{-a^+} = (\max_y |f^{(2)}(y)|)^2 = \max_y\mathcal{F}({y})$,
see FIG.~\ref{fig1}. So, the ensuing REP (\ref{tinfty1}) can be written as $N_{\infty}( \mathcal{F}) = 2^{2 a^+/D}/(2\pi) $.\\[2mm]

$b)$ When $r = -1/2$ we should work with the Fourier image $f^{(1)}$ (where $\sqrt{\hat{\mathcal{F}}} \equiv |f^{(1)}|$). This gives the
entropy power
\begin{eqnarray}
N_{1/2}(\hat{\mathcal{F}})\ = \ \frac{1}{8\pi}\exp\left(\frac{4}{D} \log|\!|f^{(1)}|\!|_{1}\right).
\label{rhalf1}
\end{eqnarray}
Results in (\ref{tinfty1})  and (\ref{rhalf1})  imply the REPUR behavior
\begin{eqnarray}
N_{1/2}(\hat{\mathcal{F}})N_{\infty}( \mathcal{F}) \ = \ \frac{1}{16\pi^2}\left(\frac{|\!|f^{(1)}|\!|_{1}}{|\!|f^{(2)}|\!|_{\infty}}  \right)^{4/D} \ \geq \ \frac{1}{16\pi^2} \ \, .
\label{repurasympt}
\end{eqnarray}
The last inequality results from the Beckner--Babenko inequality $|\!|f^{(1)}|\!|_{1} \geq |\!|f^{(2)}|\!|_{\infty} = \max_y |f^{(2)}(y)|$.
At the same time we have (consider for simplicity $D=1$)
\begin{eqnarray}
| f^{(2)}({y})| \ &=& \ \left|\int_{\mathbb{R}} d{z} \ \!  f^{(1)}({z}) e^{i2\pi yz} \right| \ = \ \left|\int_{\mathbb{R}} d{z} \ \!  |f^{(1)}({z})|
e^{i  (\theta(z) + 2\pi yz)} \right| \nonumber\\[2mm]
&=& \ \sqrt{\left( \int_{\mathbb{R}} d{z} \ \!  |f^{(1)}({z})|  \right)^2 \left\{\left[\frac{\int_{\mathbb{R}} d{z} \ \!  |f^{(1)}({z})| \cos(\theta(z) + 2\pi yz)}{\int_{\mathbb{R}} d{z}
		\ \!  |f^{(1)}({z})| }   \right]^2  + \left[\frac{\int_{\mathbb{R}} d{z} \ \!  |f^{(1)}({z})| \sin(\theta(z) + 2\pi yz)}{\int_{\mathbb{R}} d{z}
		\ \!  |f^{(1)}({z})| }   \right]^2 \right\} }\nonumber \\[2mm]
&\leq& \ \sqrt{\left( \int_{\mathbb{R}} d{z} \ \!  |f^{(1)}({z})|  \right)^2 \left\{\frac{\int_{\mathbb{R}} d{z} \ \!
		|f^{(1)}({z})| [\cos^2(\theta(z) + 2\pi yz) + \sin^2(\theta(z) + 2\pi yz)]}{\int_{\mathbb{R}} d{z}
		\ \!  |f^{(1)}({z})| } \right\}
} \ = \ |\!|f^{(1)}|\!|_{1}\, ,
\label{II.16.aac}
\end{eqnarray}
%
%
(here $\theta(z)$ is a phase of $f^{(1)}({z})$). The inequality on the third line is due to Jensen's inequality for convex functions and it saturates {\em only} when
$\theta(z) + 2\pi yz$ is a constant~{\cdred{[30]}}. This means that $\theta(z)$ must be linear in $z$, i.e., $\theta(z) = a - b z$ and since the
Beckner--Babenko inequality is saturated in this case (i.e., $|\!|f^{(1)}|\!|_{1} = |\!|f^{(2)}|\!|_{\infty}$),
we must have $b = 2\pi y_0 $ with $y_0$ representing the value of $y$ at which
$| f^{(2)}({y})|$ is maximal.
As a byproduct (\ref{II.16.aac}) boils down to
\begin{eqnarray}
\int_{\mathbb{R}} d{z} \ \! \hat{\mathcal{F}}^{1/2}({z}) \ = \ | f^{(2)}({y}_0)| \ = \ {{\mathcal{F}}^{1/2}({y}_0)} \ = \ 2^{-a^+/2}\, .
\label{II.17.a}
\end{eqnarray}
The foregoing proof applies equally well in higher-dimensional spaces. In particular, we can easily generalize our considerations to $D$ dimensions and arrive at $N_{1/2}(\hat{\mathcal{F}}) = 2^{-2 a^+/D}/(8\pi)$.

Let us now assume that the REPUR (\ref{repurasympt}) is saturated. In such a case $| f^{(2)}({y})|$ in the neighborhood of the point $y_0$ [cf. second line in (\ref{II.16.aac})] reads
\begin{eqnarray}
| f^{(2)}({y})|  &=&  |\!|f^{(1)}|\!|_{1}\sqrt{\langle \cos[2\pi (y-y_0)\mathcal{Z}] \rangle^2 + \langle \sin[2\pi (y-y_0)\mathcal{Z}] \rangle^2}\nonumber \\[1mm]
&=& |\!|f^{(1)}|\!|_{1} \sqrt{1  -  (2 \pi)^2(y-y_0)^2[\langle \mathcal{Z}^2 \rangle - \langle \mathcal{Z} \rangle^2 ] + \mathcal{O}[(y-y_0)^4]} \, .
\label{32a}
\end{eqnarray}
Here the mean $\langle \ldots \rangle$ is with respect to the PDF  $|f^{(1)}(z)|/\int_{\mathbb{R}} d{z} \ \!  |f^{(1)}({z})| $.
For small $|y-y_0|$ the equation (\ref{32a}) allows us to write
\begin{eqnarray}
{{\mathcal{F}}({y})} \ \approx \  2^{-a^+}\ \! \exp\left[-(2\pi)^2(y-y_0)^2 \tilde{\sigma}^2\right]\, ,
\label{33.a}
\end{eqnarray}
where $\tilde{\sigma}^2 \equiv \langle \mathcal{Z}^2 \rangle - \langle \mathcal{Z} \rangle^2$.
So when the REPUR (\ref{repurasympt}) is saturated then ${{\mathcal{F}}({y})}$ has a Gaussian peak and
accordingly the tails of $\hat{\mathcal{F}}({z})$ have to be Gaussian. The latter holds because the behavior of $f^{(1)}({z})$ at large values of $|z|$ is determined by the behavior of $f^{(2)}({y})$ at small values of $|y-y_0|$. Because the small $|y-y_0|$ of  $f^{(2)}({y})$  leads to a Gaussian shape, the far tails of $f^{(1)}({z})$ must be Gaussian and of the form (modulo unimportant phase factor)
\begin{eqnarray}
f^{(1)}({z}) \ \sim \   \exp\left[-\frac{z^2}{2 \tilde{\sigma}^2}\right]\,\,\, \Rightarrow \,\,\, \, \hat{\mathcal{F}}({z}) \ \sim \ \exp\left[-\frac{z^2}{\tilde{\sigma}^2}\right]\, .
\label{34.a}
\end{eqnarray}

This necessary condition is also sufficient. Namely, when the peaks and tails of ${{\mathcal{F}}({y})}$ and  $ \hat{\mathcal{F}}({z})$ are Gaussian of the form (\ref{33.a}) and (\ref{34.a}) and when, in addition, the associated ``wave function'' $f^{(1)}({z})$ has a linear phase factor then the REPUR (\ref{repurasympt}) is saturated. This can be seen by reading the inequality (\ref{II.16.aac}) backwards.

On the other hand, PDFs which do not belong to $L_{\infty}({\mathbb{R}}^D)$ (i.e., unbounded PDFs) or to $L_{1/2}({\mathbb{R}}^D)$ (i.e., heavy tailed PDFs which decay at infinity slower than
$|z|^{-{2D}}$) cannot saturate REPUR (\ref{repurasympt}) even though their respective Fourier
images can.
In fact, the corresponding REPURs are in these cases
typically undetermined (regulator dependent) as illustrated by our second quantum-mechanical system in the main text. It should be also stressed that the wave function related to $f^{(1)}$ must have a linear phase factor to allow for a saturation. The phase factor in turn determines the peak position
of the distribution ${{\mathcal{F}}({y})}$.

This anomalous asymptotic behavior  can be interpreted as follows: the non-linear nature of the RE emphasizes
the more probable parts of the PDF
(typically the middle parts) for R\'{e}nyi's index $p > 1$ while for $p < 1$ the
less probable parts of the PDF (typically the tails) are accentuated. So, when the emphasized parts
in $\mathcal{F}$ and $\hat{\mathcal{F}}$ are close to (or coincide with)  Gaussian PDF
sectors, the associated REPUR will approach the lower bound $\hbar^2/4$.
Particularly, in the asymptotic regime when $r = -1/2$, the saturation of the REPUR means that the
peak of ${\mathcal{F}}^{(1)}$ and tails of ${\mathcal{F}}^{(2)}$ are Gaussian, though both  ${\mathcal{F}}^{(1)}$ and ${\mathcal{F}}^{(2)}$
might be non-Gaussian. This fact was illustrated on both studied examples in the
main text [cf. also FIG~\ref{fig3}].\\

\textit{Acknowledgements} --- The authors thank Erin Clark for helpful comments
and critical discussions related to this Supplemental Material.

%
%